\documentclass[pra, twocolumn]{revtex4-1}

\usepackage{amsmath}
\usepackage{amsfonts}
\usepackage{amssymb}
\usepackage{multirow}
\usepackage{color}

\begin{document}
\title{Ex ante versus ex post equilibria in classical Bayesian games with a nonlocal resource}
\author{M\'aty\'as Koniorczyk}
\affiliation{Wigner Research Centre for Physics, H-1525 Budapest, P.O. box 49, Hungary}
\affiliation{Institute of Mathematics and Informatics, Faculty of Science, University of P\'ecs, H-7624 P\'ecs, Ifj\'us\'ag \'utja 6, Hungary}
\author{Andr\'as Bodor}
\affiliation{Institute of Mathematics and Informatics, Faculty of Science, University of P\'ecs, H-7624 P\'ecs, Ifj\'us\'ag \'utja 6, Hungary}
\author{Mikl\'os Pint\'er}
\affiliation{Institute of Mathematics, Budapest University of Technology and Economics, H-1111 Budapest, Egry J\'ozsef u. 1, Hungary}

\begin{abstract}
  We analyze the difference between ex ante and ex post equilibria in
  classical games played with the assistance of a nonlocal (quantum or
  no-signaling) resource. In physics, the playing of these games is
  known as performing bipartite Bell-type experiments. By analyzing
  the Clauser-Horn-Shimony-Holt game, we find a constructive procedure
  to find two-person Bayesian games with a nonlocal
  (i.e. no-signaling, and, in many cases, quantum) advantage. Most
  games of this kind known from the literature can be constructed
  along this principle, and share the property that their relevant
  ex ante equilibria are ex post equilibria as well. We introduce a
  new type of game, based on the Bell-theorem by V\'ertesi and Bene,
  which does not have the latter property: the ex ante and ex post
  equilibria differ.
\end{abstract}
\maketitle

\section{Introduction}

It is widely accepted that Bell-type experiments and nonlocality are
deeply related to game theory. It is known that in a classical
strategic game the use of nonlocal signals may have certain benefits,
such as new, better Pareto equilibria (see
e.g. Ref.~\cite{LaMura2005}). Brunner and Linden~\cite{Brunner2013}
have pointed out the deep relation between Bell-inequalities and the
theory of Bayesian games leading to a new insight in case of both
problems. This line of research is not to be confused with ``quantum
game theory''~\cite{Eisert1999}, which is dealing with scenarios
constructed from classical non-Bayesian games by a specific
quantization procedure and not involving quantum or other nonclassical
correlations. That kind of quantization of a classical game leads to a
modified classical game~\cite{vanEnk2002}, while in our case the
classical game is Bayesian and not quantized, but it is played with
the aid of a nonlocal resource possibly implemented with quantum
systems.

Nonlocal correlations can lead to new equilibria in classical Bayesian
games. This has attracted significant research attention.  The
general notion of these equilibria was studied by Auletta et
al.~\cite{Auletta2017}. Brandenburger and La
Mura~\cite{Brandenburger2016} addressed the question of quantum
improvement from the perspective of team decision problems. Meanwhile,
there are many particular classical games known for having an
improvement when quantum nonlocal resources can be used. Examples
include the ``prototype'' CHSH game which is a game-type description
of the Bell-CHSH inequality~\cite{PhysRevLett.23.880}, the
entanglement-assisted orientation in space by
Brukner~\cite{BRUKNER2006}, or the high-frequency trading example
described in Ref.~\cite{Brandenburger2016} just to mention a few. The
CHSH game also has a conflict of interest variant~\cite{Pappa2015}
which inspired detailed studies of games with a similar equilibrium
structure~\cite{Rai2017,Roy2016,Bolonek-Lason2017}.

These games are \emph{ad-hoc} constructions; a general mathematical
recipe leading systematically to similar games is yet to be explored,
and so is a systematic approach to find a game which benefits from the
violation of a known Bell-inequality. In the present contribution we
deal with two-player Bayesian games having a quantum advantage in
order to make steps towards this direction. The detailed examination
of the CHSH game leads us to a generic construction of games of the
above kind which covers the other games, and it can serve as a
recipe to find additional games of this kind. This common route of
many games with a quantum advantage has not been pointed out before.

Having a Bell-type inequality at hand, it can be instructive to
examine how its left-hand side relates to our construction. Searching
for the maximal value of the left-hand side of the Bell inequality
corresponds to a situation termed as \emph{ex ante} in Bayesian game
theory, whereas our construction originates from the \emph{ex post}
situation. These two eventually coincide in the case of the games
considered in the literature. Consequently, the difference between
these has not yet been explored.

However, the two situations can indeed differ, thereby leading to
games with a different equilibrium structure. Here we introduce such a
game: studying the Bell-type inequality by V\'ertesi and
Bene~\cite{Vertesi2010}, we find one which does not follow our
construction and thus has an equilibrium structure different from that
of the other known games.

This paper is organized as follows. In Section~\ref{sect:gametheor} we
summarize some concepts of game theory we shall
use. Section~\ref{sect:Bayes} is devoted to concepts related to
Bayesian games, and the relation of those games to Bell-type
experiments. On this basis we describe the structure of many
Bayesian games related to Bell-type experiments from the point of view
of their equilibria, and we describe a means for constructively
defining such games. In Section~\ref{sect:VB} we study the
V\'ertesi-Bene Bell-inequality to show that it leads to a game with a
different equilibrium structure. In Section~\ref{sect:conclusions} our
results are summarized and conclusions are drawn.

\section{Game theory overview}
\label{sect:gametheor}

Let us start with a brief review of some concepts of game theory; we
refer to Ref.~\cite{gtbook} for their detailed description. Games in
strategic form are common models in game theory. In case of two
players and finite strategy sets, they are also termed as bimatrix
games. A bimatrix game is a quadruple
$(A, B, u_{\text{A}},u_{\text{B}})$, where $A$ and $B$ are finite sets
representing the set of actions of the two players, Alice and Bob,
respectively, whereas $u_{\text{A}}$ and $u_{\text{B}}$ are
$A\times B \mapsto \mathbb{R}$ functions representing the payoff (or
utility) of each player. Throughout this paper we shall, without loss
of generality, label the actions with numbers.

A central concept of game theory is that of the equilibrium. The
(pure) Nash equilibrium of a game~\cite{Nash1950} is a selection of
actions so that it is not worth deviating from the given strategy by
any of the players unilaterally: $(a,b)$ is a Nash equilibrium if
\begin{eqnarray}
  \label{eq:pure_equil}
  \left(\forall a'\in A\right) \quad u_{\text{A}}(a',b) \leq u_{\text{A}}(a,b),\nonumber \\
  \left(\forall b'\in B\right) \quad u_{\text{B}}(a,b') \leq u_{\text{B}}(a,b).
\end{eqnarray}

As an example consider the game called ``battle of sexes'', defined in
the left of Tab.~\ref{tab:bs}.
\begin{table}
  \centering
  $$
  \begin{array}{|r||c|c|}
    \hline
    a\downarrow b\rightarrow& 0  & 1  \\
    \hline 
    \hline
    0  & (2,1) & (0,0)  \\
    1  & (0,0) & (1,2)  \\
    \hline
  \end{array}\qquad
    \begin{array}{|r||c|c|}
    \hline
    a\downarrow b\rightarrow& 0  & 1  \\
    \hline 
    \hline
    0  & (1,1) & (0,0)  \\
    1  & (0,0) & (1,1)  \\
    \hline
    \end{array}
    $$
    \caption{The Battle of sexes game (left) and the coordination game
      (right). The columns and rows refer to the players' actions
      $a\in A$, $b\in B$, whereas the numbers in the tables are the
      payoffs in the format $(u_\text{A}(a,b),u_\text{B}(a,b))$.}
\label{tab:bs}
\end{table}
(The common story behind it is that Alice prefers going to the cinema
whereas Bob prefers going to a soccer match, but they would
like to be together.) The game has two pure equilibria, $a=b=0$ and
$a=b=1$.

An extension to the concept of strategies is to consider probability
distributions over the strategies, so-called mixed strategies. Having
$p_{\text{A}}(a)$ and $p_{\text{B}}(b)$, two independent probability
distributions on Alice's and Bob's side, one can calculate the
expected value of the utility, the \emph{expected payoff} using the
joint distribution being the product of those on the two sides. The
probability distributions are termed as \emph{mixed strategies} as
opposed to the deterministic \emph{pure strategies} mentioned so
far. The concept of the equilibrium can be extended in a
straightforward manner by considering mixed strategies and the
expected payoff. According to the celebrated theorem by
Nash~\cite{Nash1950,Nash1951}, a mixed equilibrium always exists in
the games studied here. In the case of the ``battle of sexes'', in
addition to the aforementioned pure equilibria, there is also a mixed
one with the probability distributions $p_{A}=(2/3, 1/3)$ and
$p_{B}=(1/3, 2/3)$, resulting in the expected payoff of $(2/3, 2/3)$.

Finally, an even more general equilibrium concept is that of
correlated equilibrium~\cite{Aumann1974}, in which we allow for a
joint probability distribution $p(a,b)$ to determine the choice of the
players. Its definition reads
\begin{eqnarray}
  \label{eq:eq_def}
  \left(\forall a,a'\right)\ \sum_{b} (u_{\text{A}}(a,b)-u_{\text{A}}(a',b))p(a,b) \geq 0 \nonumber \\
  \left(\forall b,b'\right)\ \sum_{a} (u_{\text{B}}(a,b)-u_{\text{B}}(a,b'))p(a,b) \geq 0.
\end{eqnarray}
From amongst the mentioned equilibrium concepts, this is maybe the
easiest to tackle mathematically as the equilibrium strategies form a
polytope in the set of probability distributions. If, however, this
distribution happens to be correlated, then in order to implement the
strategy the players may have to rely on the advice of a trusted
coordinator or they have to agree in advance on the strategies. An
example of such an equilibrium appears e.g. in the ``coordination
game'', defined in the right of Tab.~\ref{tab:bs}. Here
$p(0,0)=p(1,1)=1/2$ is a correlated equilibrium (there are infinitely
many other correlated equilibria though). Coordination games
frequently appear in quantum-assisted situations. The equilibrium in
the correlated case takes the intuitive meaning that it is not worth
deviating from the advice of the trusted coordinator
unilaterally. 

\section{Bayesian games and Bell-type experiments}
\label{sect:Bayes}

It is prevalently accepted that a Bell-type experiment can also be
interpreted as a kind of game termed as ``Bayesian game'' in the
language of game theory. In such games, Alice and Bob choose some
input first (or get it from a third party). In physics we say that
both Alice and Bob choose a \emph{measurement}, whereas in the game
theory language, they become aware of their \emph{types} $x\in X$ and
$y\in Y$ first ($X$ and $Y$ are finite sets). Then they choose an
action, and the payoff depends on both the types and the actions, so
we have $u_{\text{A}}(a,b,x,y)$ and $u_{\text{B}}(a,b,x,y)$,
respectively, thus we have a strategic form game for each pair of
types $(x,y)$.  (Note that in many applications of Bayesian games the
payoff does not depend on the other player's type.) In the Bell-type
experimental scenario the actions are the results $a\in A$ ($b\in B$)
of the measurement $x\in X$ ($y\in Y$) the latter playing the role of
the types. They are assumed to be chosen by the parties independently
at will; this choice can thus be modeled as a probability distribution
of their possible types (aka measurements). In order to verify
e.g. the violation of local realism, some expression of the outcomes
(aka actions) is evaluated.

As of the equilibria of Bayesian games, there are three situations
studied in game theory. In the \emph{ex ante} situation, no players
know each others' types and neither that of their own, thus regarding
equilibria, the relevant payoffs are expected values on some prior
distribution of all types. The \emph{interim} case is when each player
knows his own type but has only a probability distribution on that of
the others'. Hence, when the payoff is studied, each player considers
the expected value on the others' types. It is prevalently known in
game theory that in the case of the games studied here the set of ex
ante and interim equilibria coincide. Finally, in an \emph{ex post}
situation, all players know everyone's types, thus we fall back to the
case of simple strategic form games. So for a given $(x,y)$, we can
consider the equilibrium properties of the game defined by the actions
and the respective payoffs $u_{\text{A}}(a,b,x,y)$,
$u_{\text{A}}(a,b,x,y)$ with $x$ and $y$ fixed. Hence, the set of
strategies are classified by the conditional probability distributions
$P(a,b|x,y)$ which describe the \emph{behavior} of the nonlocal
resource in a Bell-type scenario.

Consider, for instance, the case of the CHSH experiment (CHSH
game). The payoff is the same for both players:
$u_{\text{A}}=u_{\text{B}}=u$; it is tabulated on the upper part of
Tab~\ref{tab:chsh}.
\begin{table}
  \centering
  $$
  \begin{array}{|lll||cc|cc|}
    \hline
    x\downarrow& y\rightarrow&&\multicolumn{2}{|c}{0}&  \multicolumn{2}{c|}{1}\\
    &a\downarrow& b\rightarrow& 0  & 1 &  0  & 1  \\
    \hline 
    \hline
    \multirow{2}{*}{0}&0  & &  1 & -1  & 1  & -1    \\
    \cline{2-3}&1&&            -1  & 1 & -1  & 1   \\
    \hline
    \multirow{2}{*}{1}&0  & &  1  & -1  & -1 & 1   \\
    \cline{2-3}&1&&              -1  & 1  & 1  & -1  \\
    \hline
  \end{array}
  $$
  $$
  \begin{array}{|lll||cc|cc|}
    \hline
    x\downarrow& y\rightarrow&&\multicolumn{2}{|c}{0}&  \multicolumn{2}{c|}{1}\\
    &a\downarrow& b\rightarrow& 0  & 1 &  0  & 1  \\
    \hline 
    \hline
    \multirow{2}{*}{0}&0  & &  1/2 & 0 & 1/2  & 0    \\
    \cline{2-3}&1&&            0 & 1/2 & 0  & 1/2   \\
    \hline
    \multirow{2}{*}{1}&0  & &  1/2 & 0 & 0 & 1/2    \\
    \cline{2-3}&1&&            0  & 1/2  & 1/2  & 0  \\
    \hline
  \end{array}
  $$
\caption{The payoff function of the CHSH game (top) and the PR-box, the
no-signaling behavior maximizing it (bottom). The tabular divides into blocks
indexed with types. Within each pair of types, we have a game payoff
matrix similar to that in~\ref{tab:bs}. There is a single
payoff value in each entry as the payoffs are equal for both
parties.}
\label{tab:chsh}
\end{table}

The expected payoff for such a game, given a uniform prior on the
types $(x,y)$ reads
\begin{equation}
  \label{eq:BayesLHS}
  U=\frac14 \sum_{x,y}\sum_{a,b} u(a,b,x,y)P(a,b|x,y),
\end{equation}
thus it is linear in the conditional probability characterizing the
behavior. Based on the assumption on the behaviors as constraints of
its maximization, it can attain various maxima.

A possible assumption is locality: the conditional probabilities are
the convex combinations of products of deterministic local channels
(i.e. which assign a set of pure strategies to each type). Hence, they
form a polyhedron in the vector space of conditional probabilities
$P(a,b|x,y)$, which is thus naturally given in its vertex
representation, that is, as a convex combination of its aforementioned
vertices~\cite{PhysRevA.78.032116,RevModPhys.86.419}. This polytope is
termed as the \emph{local polytope} $\mathcal{P}_{\text{L}}$. The
maximization of the payoff in \eqref{eq:BayesLHS} subject to
$P(a,b|x,y)\in \mathcal{P}_{\text{L}}$ is a linear program. The CHSH
inequality's RHS is this local maximum, which is 1/2 in this
case. (Recall that we calculate the ex ante expected payoffs with a
uniform prior on the types. In the classic literature of the Bell-CHSH
inequality, the prevalently known local bound is 2, which is the
ex-post value, hence our bounds are to be multiplied by 4 to achieve
the usual values.)

A broader class of behaviors is that of the quantum ones. The set of
quantum behaviors in the space of conditional probabilities is a
convex domain which is a superset of $\mathcal{P}_{\text{L}}$, and can
be characterized by an infinite series of semidefinite
programs~\cite{PhysRevLett.98.010401, 1367-2630-10-7-073013}. The
heuristically determined maximal value is $\sqrt 2/2$ known as
Tsirelson's bound~\cite{Cirelson1980}.

Finally, the broadest set frequently considered is that of the
no-signaling behaviors~\cite{RevModPhys.86.419}. In their case the
single assumption is that the given resource cannot be used for
transmitting any classical information between the two
parties. Mathematically this can be formulated as the
no-signaling conditions:
\begin{eqnarray} \forall a, x, y_1, y_2 \quad \sum_b P (a, b | x, y_1 ) =_{}
   \sum_b P (a, b | x, y_2 ) ,\nonumber \\
\forall b, x_1, x_2, y_{} \quad \sum_a P (a, b | x_1, y_{} ) =_{}
   \sum_a P (a, b | x_2, y_{} ), \label{nosig2} 
\end{eqnarray}
defining a polytope $\mathcal{P}_{\text{NS}}$ in the space of
conditional probabilities again.  Solving the linear program of
finding the maximal payoff in \eqref{eq:BayesLHS} subject to
$P(a,b|x,y)\in \mathcal{P}_{\text{NS}}$ leads to the objective value
of $1$. (This corresponds to the ex-post value of 4 which is known in the
Bell-inequality literature.) As in case of all such linear programs,
it is obtained at a vertex of $\mathcal{P}_{\text{NS}}$, outside
$\mathcal{P}_{\text{L}}$ and also outside the quantum domain. We
tabulate this behavior in the lower part of Tab.~\ref{tab:chsh}.  The
fictitious device realizing this is the Popescu-Rorhlich (PR)
box~\cite{Rastall1985,KT85,Popescu1992,Popescu1994}.

Let us now make two important observations. Firstly, as the payoffs in
these games are equal, maximizing the expected payoff necessarily
leads to an ex ante equilibrium. Were this not the case, it would be
possible for at least one player to unilaterally deviate, thereby
obtaining a better payoff. As the payoff of the other party is the
same, a better payoff could be obtained which would be higher than the
maximum, which is impossible.  Secondly, recall that if once
$(x_0,y_0)$ is fixed, in the ex post situation the payoff depends only
on the actions, and hence the block $P(a,b|x_0,y_0)$ of the
conditional distribution becomes a (possibly correlated) strategy
profile of the game corresponding to $(x_0,y_0)$. Hence, given such a
distribution it can be verified whether it is an equilibrium using the
definitions of equilibrium of bimatrix games. \emph{In the case of the
  CHSH game, given any particular pair of types $(x,y)$, the
  respective block of the optimal no-signaling behavior is an
  equilibrium of the game.} Otherwise speaking, an ex post equilibrium
is achieved. A possible intuitive interpretation is the following. A
crucial assumption of the Bell-situations is that none of the parties
is aware of the other's type. And indeed, we are using a PR-box which
is a no-signaling resource. However, our observation implies that it
is not worth for any of the players to deviate from the advice of the
box, even if they would become aware of the other's type which is the
ex post situation. A direct calculation shows that the same holds for
the optimal quantum behavior, as it is a convex combination of the
PR-box and the uniform distribution in all blocks.

Studying further the construction of the CHSH game another observation
can be made. If we were to search for a game which is interesting from
the point of view of nonlocal resources and we wouldn't be aware of
this one, it would be possible to construct it as follows. Pick a
nonlocal vertex of the no-signaling polytope
$\mathcal{P}_{\text{NS}}$, this defines a behavior. Then define the
payoff matrix of a Bayesian game so that for all the blocks for given
types $(x,y)$, the corresponding probability distribution is its
correlated equilibrium according to Eq.~\eqref{eq:eq_def}. For a fixed
behavior $p(a,b)$, Eq.~\eqref{eq:eq_def} defines a nonempty polyhedron
in the space of payoff matrices, so there may be infinitely many
choices. In addition, the payoff should be chosen such a way that the
maximum of the expected payoff for a selected prior (e.g. the uniform
distribution of types) over the local polytope
$\mathcal{P}_{\text{L}}$ is lower than the no-signaling value on the
chosen vertex. In our particular example, the CHSH game could be
constructed entirely from the vertex: the PR-box, which is the only
nontrivial nonlocal no-signaling vertex in the two-input-two-output
case. The CHSH game can be obtained along the above lines.  As the
expected payoff is a linear function, and the quantum domain is a
convex one between $\mathcal{P}_{\text{L}}$ and
$\mathcal{P}_{\text{NS}}$, it is likely that there will be a quantum
advantage, too.

All the two-player games with a quantum advantage we are aware of
derive from this construction. The examples include the high-speed
trading example by Brandenburger and La Mura~\cite{Brandenburger2016}
and the spatial orientation game by Bruckner~\cite{BRUKNER2006}, two
games found in a heuristic way independently from each other. These
two different payoffs can be derived from the same vertex. The
``conflict of interest game'' in Ref.~\cite{Pappa2015}, even though it
is not directly a coordination game, also obeys this kind of
construction, using the same vertex as the CHSH game.

\section{An exception: the V\'ertesi-Bene game}
\label{sect:VB}

The question naturally arises whether there are examples of Bayesian
games with a nonlocal advantage which do not obey our construction,
and have a different equilibrium structure.  In search for such an
exception, let us investigate the bipartite Bell-situation which was
studied by V\'ertesi and Bene in Ref.~\cite{Vertesi2010} which was
derived to prove the possible nontrivial role of POVM measurements in
Bell-type scenarios.  A peculiar feature of this case is that Alice
has 3 inputs, and only for the third input there are 3 outputs. In
case of the first two inputs, the number of outputs is 2. Bob has two
inputs and two outputs for both.
The payoff function is tabulated in Table~\ref{tab:ICH3}, with a
parameter $c>0$.
\begin{table}
  $$
  \begin{array}{|lll||cc|cc|}
    \hline
    x\downarrow& y\rightarrow&&\multicolumn{2}{|c}{0}&  \multicolumn{2}{c|}{1}\\
    &a\downarrow& b\rightarrow& 0  & 1 &  0  & 1  \\
    \hline 
    \hline
    \multirow{2}{*}{0}&0  & &   0   & -c/2 & c/2 & -c/2 \\
    \cline{2-3}&1&&             -c/2 &  0   &  0  &  0   \\
    \hline
    \multirow{2}{*}{1}&0  & &   c/2  &  0   & -c  &  0   \\
    \cline{2-3}&1&&            -c/2 &  0   &  0  &  0   \\
    \hline
    \multirow{3}{*}{2}&0  & &  1/2   &  -1/2   &  1/2  &  -1/2   \\
    \cline{2-3}&1&&              \frac{\sqrt{2}+2}{4}   &  \frac{\sqrt{2}-2}{4}   &  \frac{\sqrt{2}-6}{4}  &  \frac{\sqrt{2}-2}{4}   \\
    \cline{2-3}&2&&              0   &  0   &  0  &  0   \\
    \hline
  \end{array}
  $$
  \caption{The payoff of the V\'ertesi-Bene game in Ref.~\cite{Vertesi2010}, $c>0$ is a parameter.}
  \label{tab:ICH3}
\end{table}
This attains a local bound of $1$, e.g. by the vertex of the local
poytope tabulated in Table~\ref{tab:ICH3locmax}, apparently regardless
of the value of $c$. Note that the blocks typeset in bold are not
equilibria of the respective game (the bimatrix one defined by the
same block of the payoff), thus clearly the local vertex maximizing
the expected payoff belongs to a non-equilibrium behavior.
\begin{table}
  $$
  \begin{array}{|lll||cc|cc|}
    \hline
    x\downarrow& y\rightarrow&&\multicolumn{2}{|c}{0}&  \multicolumn{2}{c|}{1}\\ 
    &a\downarrow& b\rightarrow& 0  & 1 &  0  & 1  \\
    \hline 
    \hline
    \multirow{2}{*}{0}&0  & &   1   & 0 & 1 & 0 \\
    \cline{2-3}&1&&             0 &  0   &  0  &  0   \\
    \hline
    \multirow{2}{*}{1}&0  & &   1  &  0   & \mathbf 1  &  \mathbf 0   \\
    \cline{2-3}&1&&             0 &   0   &  \mathbf 0  &  \mathbf 0   \\
    \hline
    \multirow{3}{*}{2}&0  & &  \mathbf 1   &  \mathbf 0   &  1  &  0   \\
    \cline{2-3}&1&&            \mathbf 0   &  \mathbf 0   &  0 &  0   \\
    \cline{2-3}&2&&            \mathbf  0   & \mathbf 0   &   0  &  0   \\
    \hline
  \end{array}
  $$
  \caption{An optimal local behavior for the game in
    Table~\ref{tab:ICH3}. The blocks in boldface are not
    equilibria of the game in the corresponding block.}
  \label{tab:ICH3locmax}
\end{table}

Let us now leave the local polytope and check the no-signaling
case. It is the linear programming problem of finding the maximum of
average payoff assuming an uniform prior on the types. This is the
objective function, and the feasibility region is the local polytope
$\mathcal{P}_{\text{L}}$, defined by the trivial normalization of the
behaviors and the no-signaling conditions in Eq.~\eqref{nosig2}. For
$c>1$, the optimum is achieved at a vertex which corresponds to the
nonlocal behavior tabulated in Tab.~\ref{tab:vbns}.  Observe that the
optimal no-signaling vertex has a block (in particular, the one
belonging to Alice's third type and Bob's first type), which is not an
equilibrium of the game.  \emph{So in this case the ex ante and ex
  post equilibria differ.} In the case of other games like the CHSH
the presence of an optimal no-signaling box can assist in the
realization of an equilibrium which could not be further improved by
letting the players to have full information of the others' type. In
contrast to that, in the just studied game the players could further
benefit from actual information on the other's type, albeit the
nonlocal box also leads to an improvement over the classical
result. \emph{Hence, there are games that could be played better if
  the players wouldn't be restricted to no-signaling resources but
  were aware of the other's type.}  Otherwise speaking, if Alice and
Bob knew each other's type with certainty (e.g. via some communication
channel), they would play differently from having just the
no-signaling box at hand. This is a feature not present in the other
known two-player games.
\begin{table}
  \centering
  $$
  \begin{array}{|lll||cc|cc|}
    \hline
    x\downarrow& y\rightarrow&&\multicolumn{2}{|c}{0}&  \multicolumn{2}{c|}{1}\\
    &a\downarrow& b\rightarrow& 0  & 1 &  0  & 1  \\
    \hline 
    \hline
    \multirow{2}{*}{0}&0  & &   1/2   & 0 & 1/2 & 0 \\
    \cline{2-3}&1&&             0 &  1/2   &  0  &  1/2   \\
    \hline
    \multirow{2}{*}{1}&0  & &   1/2  &  0   & 0  &  1/2   \\
    \cline{2-3}&1&&            0 &  1/2   &  1/2  &  0   \\
    \hline
    \multirow{3}{*}{2}&0  & &  \mathbf{1/2}   &  \mathbf 0   &  1/2  &  0   \\
    \cline{2-3}&1&&            \mathbf 0   &  \mathbf 0   &  0 &  0   \\
    \cline{2-3}&2&&            \mathbf  0   & \mathbf{1/2}   &  0  &  1/2   \\
    \hline
  \end{array}
    $$
    \caption{The optimal no-signaling behavior for the game in
      Tab.~\ref{tab:ICH3}, for
      $c>1$. The block in boldface is not an equilibrium of the game
      in the corresponding block.}
\label{tab:vbns}
\end{table}

The question naturally arises whether this game has similar features
when a quantum box is used by the players. Ref.~\cite{Vertesi2010}
contains an example of a quantum behavior violating the classical
bound in this case (not an extremal one though). This is tabulated in
Tab.~\ref{tab:vbq}.
\begin{table}
  $$
  \begin{array}{|lll||cc|cc|}
    \hline
    x\downarrow& y\rightarrow&&\multicolumn{2}{|c}{0}&  \multicolumn{2}{c|}{1}\\
    &a\downarrow& b\rightarrow& 0  & 1 &  0  & 1  \\
    \hline 
    \hline
    \multirow{2}{*}{0}&0  & &  \frac{2+\sqrt{2}}{8} & \frac{2-\sqrt{2}}{8} & \frac{2+\sqrt{2}}{8}  & \frac{2-\sqrt{2}}{8}    \\
    \cline{2-3}&1&&            \frac{2-\sqrt{2}}{8} & \frac{2+\sqrt{2}}{8} & \frac{2-\sqrt{2}}{8}  & \frac{2+\sqrt{2}}{8}   \\
    \hline
    \multirow{2}{*}{1}&0  & &  \frac{2+\sqrt{2}}{8} & \frac{2-\sqrt{2}}{8} & \frac{2-\sqrt{2}}{8} & \frac{2+\sqrt{2}}{8}    \\
    \cline{2-3}&1&&            \frac{2-\sqrt{2}}{8}  & \frac{2+\sqrt{2}}{8}  & \frac{2+\sqrt{2}}{8}  & \frac{2-\sqrt{2}}{8}  \\
    \hline
    \multirow{3}{*}{2}&0  & &  \mathbf{0.30602}   &  \mathbf{0.12925}  &  \mathbf{0.41652}  &  \mathbf{0.01875}   \\
    \cline{2-3}&1&&            \mathbf{0.18243}   &  \mathbf{0.11444}  &  \mathbf{0.00395} &   \mathbf{0.29293}   \\
    \cline{2-3}&2&&            \mathbf{0.01155}   &  \mathbf{0.25630}  &  \mathbf{0.07953}  &  \mathbf{0.18832}   \\
    \hline
  \end{array}
  $$
  \caption{The quantum behavior with an advantage discussed in
    Ref.~\cite{Vertesi2010}. The values have been obtained by the direct 
    evaluatinon of the result in the cited paper.
    The blocks in boldface are not
    equilibria of the game in the corresponding block of
    Tab~\ref{tab:ICH3}.}
\label{tab:vbq}
\end{table}
In the case of the first two types of Alice, the CHSH case is repeated
by the construction in Ref.~\cite{Vertesi2010}, whereas in the lowest
two blocks we have evaluated their numerical
construction. Interestingly, now both of the blocks for Alice's type
$2$ (the bold-typed ones) are non-equilibria, thus this behavior is
not an \emph{ex post} equilibrium. It can be verified via direct
calculation that it is an \emph{ex ante} equilibrium meanwhile.

\section{Conclusions}
\label{sect:conclusions}

In conclusion, we have studied two-player classical Bayesian games
played with the assistance of a nonlocal resource. These correspond to
bipartite Bell-type experiments in physics. We have addressed the
interpretation of these situations in different phases of the game,
depending on what is revealed for the players. We have found that the
ex post equilibria and the ex ante ones achievable with no-signaling
or quantum resources coincide in the games studied in the
literature. In contrast to that, we have found a game in which these
equilibria are different.

It would be interesting to find some intuitive explanation of this
unusual behavior of the game. It is likely that it relates to the
high asymmetry of the given situation. This also reduces the
symmetries of the relevant polytopes, which is necessary for the
corresponding Bell theorem to prove that POVM-s can have a nontrivial
role in the theory of quantum information. What we found here is that
this Bell-situation, in addition to accomplishing this goal, also
exhibits an interesting game-theoretic structure, which can be a
subject of further research.

Refs.~\cite{Auletta2017,LaMura2005} define the notion of a variety of
equilibrium concepts of Bayesian games played in the presence of local
resources, thereby paving the way to the study of the equilibrium
structure of these games. In the present paper we find a generic
construction for these kinds of games, and reveal a particular
property of the equilibria which is not observable in every game.

\begin{acknowledgements}
  The authors acknowledge the support of the National Research,
  Development and Innovation Office (NKFIH) under the contracts
  Nos. K124631 and K124351. M.P. thanks GAMENET for support. We thank
  Gernot Alber and Tam\'as Kiss for useful
  discussions. A. B. wants to express special thanks to Tam\'as Geszti.
\end{acknowledgements}

%

\end{document}